\documentclass[a4paper,twocolumn,american,english,prl,superscriptaddress,longbibliography,fixfloat,notitlepage]{revtex4-2}
\pdfoutput=1
\usepackage[T1]{fontenc}
\usepackage[latin9]{inputenc}
\usepackage{xcolor}
\usepackage{pdfcolmk}
\usepackage{amsmath}
\usepackage{amssymb}
\usepackage{graphicx}
\PassOptionsToPackage{normalem}{ulem}
\usepackage{ulem}

\makeatletter

\pdfpageheight\paperheight
\pdfpagewidth\paperwidth

\providecolor{lyxadded}{rgb}{0,0,1}
\providecolor{lyxdeleted}{rgb}{1,0,0}
\DeclareRobustCommand{\lyxsout}[1]{\ifx\\#1\else\sout{#1}\fi}

\usepackage{xcolor}
\definecolor{darkblue}{rgb}{0.1,0.2,0.6} 
\definecolor{lightblue}{rgb}{0.1,0.1,1.0}
\definecolor{darkred}{rgb}{0.8,0.1,0.2}
\usepackage[colorlinks,citecolor=lightblue,linkcolor=darkblue,urlcolor=lightblue] {hyperref}

\makeatother

\usepackage{babel}
\begin{document}
\global\long\def\E{\mathrm{e}}%
\global\long\def\D{\mathrm{d}}%
\global\long\def\I{\mathrm{i}}%
\global\long\def\ket#1{\left|#1\right\rangle }%

\global\long\def\etal{\textit{et al.}}%
\global\long\def\tr{\text{Tr}\,}%
 
\global\long\def\im{\text{Im}\,}%
 
\global\long\def\re{\text{Re}\,}%
 
\global\long\def\bra#1{\left\langle #1\right|}%
 
\global\long\def\braket#1#2{\left.\left\langle #1\right|#2\right\rangle }%
 
\global\long\def\obracket#1#2#3{\left\langle #1\right|#2\left|#3\right\rangle }%
 
\global\long\def\proj#1#2{\left.\left.\left|#1\right\rangle \right\langle #2\right|}%
\global\long\def\mds#1{\mathds{#1}}%

\title{Chaos enhancement in large-spin chains}
\author{\selectlanguage{american}%
Yael Lebel}
\affiliation{\selectlanguage{american}%
Department of Physics, Ben-Gurion University of the Negev, Beer-Sheva
84105, Israel}
\author{\selectlanguage{american}%
Lea F. Santos}
\affiliation{\selectlanguage{american}%
Department of Physics, Yeshiva University, New York, New York 10016,
USA}
\author{\selectlanguage{american}%
Yevgeny Bar Lev}
\affiliation{\selectlanguage{american}%
Department of Physics, Ben-Gurion University of the Negev, Beer-Sheva
84105, Israel}
\email{ybarlev@bgu.ac.il}

\begin{abstract}
We study the chaotic properties of a large-spin XXZ chain with onsite
disorder and a small number of excitations above the fully polarized
state. We show that while the classical limit, which is reached for
large spins, is chaotic, enlarging the spin suppresses quantum chaos
features. We attribute this suppression to the occurrence of large
and slow clusters of onsite excitations, and propose a way to facilitate
their fragmentation by introducing additional decay channels. We numerically
verify that the introduction of such relaxation channels restores
chaoticity for large spins, so that only three excitations are required
to achieve strong level repulsion and ergodic eigenstates.
\end{abstract}
\maketitle

\section{Introduction}

Partially motivated by the Bohigas-Giannoni-Schmit conjecture \citep{Casati1980,Bohigas1984},
quantum chaos was extensively studied in the 1980's and 1990's \citep{Stockmann2007,Haake2001,Reichl1994}.
In the last ten years, the subject has seen a resurgence of interest
due to its strong connection with several questions currently studied
experimentally and theoretically, that include the issue of thermalization
in isolated many-body quantum systems \citep{Zelevinsky1996,Borgonovi2016,Alessio2016},
the problem of heating in driven systems \citep{DAlessio2014,Lazarides2014},
the difficulty to achieve many-body localization \citep{Nandkishore2014,Abanin2017,Agarwal2016_review,Luitz2016},
and the fast scrambling of quantum information \citep{Swingle2016a,Garttner2017,Borgonovi2019:exponentially,Sanchez2020,Niknam2020}.
For systems with clear classical or semi-classical limits, quantum
chaos refers to signatures found in the quantum domain, such as level
statistics as in full random matrices \citep{Guhr1998}, that indicate
whether the classical system is chaotic in the sense of positive Lyapunov
exponent and mixing. While this correspondence holds well for some
systems with a small number of degrees of freedom, such as Sinai's
billiard \citep{Casati1980,Bohigas1984}, it has recently been shown
to be violated in triangular billiards \citep{Lozej2021} and quantum
triangle maps \citep{Wang2022}. As one moves to systems with many
interacting particles, this issue gets even more complicated, since
the classical limit is not always straightforward \citep{PhysRevLett.118.164101}.

In this work, we investigate a one-dimensional system of many interacting
spins described by the Heisenberg XXZ model with nearest-neighbor
couplings and onsite disorder. We employ the term ``quantum chaos''
as a synonym for level statistics as in random matrices. For the \$z\$-direction
magnetization close to zero, this model is chaotic for spin-1/2 \citep{Avishai2002},
spin 1 \citep{Richter2019,Santos2020} , and larger spins \citep{deWijn2012,Elsayed2015}.
For spin-1/2, the model has also been shown to demonstrate chaotic
traits for as little as 3 or 4 excitations above a fully polarized
state of spins \citep{schiulaz2018:FewManybodyQuantum,Zisling2021}
and even for a chain of only 3 spins-1/2 \citep{Mirkin2021:QuantumChaos}.
Here, we extend this analysis and examine the case of large-spin chains
with a low number of excitations. In the semi-classical limit of a
continuous spin, we verify that the system has a positive Lyapunov
exponent. This might suggest that increasing the spin size would require
even fewer excitations than in the case of spin-1/2 to reach the quantum
chaotic regime. However, rather counterintuitively, the opposite takes
place, larger spin takes us away from quantum chaos which we attribute
to the emergence of clusters of onsite excitations. Therefore to achieve
quantum chaos in the zero-density excitations limit, we need to modify
the Hamiltonian by adding terms that ensures the fragmentation of
those clusters. By doing so, quantum chaos can finally be reached
with only 3 excitations.

\section{The model\label{sec:Large-Spin-XXZ}}

We consider the large-spin version of the XXZ model with onsite disorder
and open boundaries described by the following Hamiltonian

\begin{eqnarray}
\hat{H} & = & \frac{J_{xy}}{s\left(s+1\right)}\sum_{k=1}^{L-1}\left(\hat{S}_{x}^{k}\hat{S}_{x}^{k+1}+\hat{S}_{y}^{k}\hat{S}_{y}^{k+1}\right)\label{eq:XXZ model}\\
 &  & +\frac{J_{z}}{s\left(s+1\right)}\sum_{k=1}^{L-1}\hat{S}_{z}^{k}\hat{S}_{z}^{k+1}+\frac{1}{\sqrt{s\left(s+1\right)}}\sum_{i=1}^{L}h_{k}\hat{S}_{z}^{k},\nonumber 
\end{eqnarray}
where $\hat{S}_{\alpha}^{k}$ , with $\alpha=x,y,z$, stands for spin-$s$
operators acting on a lattice site $k$ with eigenvalues $S_{\alpha}^{k}\in\left[-s,s\right]$.
The parameter $J_{xy}$ corresponds to the coupling strength in the
$xy$ plane and $J_{z}$ stands for the strength of the interaction
along the $z$-axis. To stay away from the isotropic point, $J_{xy}=J_{z}$,
we choose $J_{xy}=1$ and $J_{z}=0.55$. The onsite disorder, where
$h_{k}$ is independent and uniformly distributed random numbers in
the interval $\left[-W,W\right]$, is introduced to break spatial
symmetries. We use a weak amplitude, $W=0.5$, to avoid possible localization
effects. The model conserves the total $z-$magnetization $\sum_{k}\hat{S}_{k}^{z}$.

Since we are interested in the large spin limit of the model, the
couplings and disorder parameters are normalized, such that all terms
in the Hamiltonian have the same magnitude. For spin-$1/2$ and $W\sim J_{xy}\sim J_{z}$,
model (\ref{eq:XXZ model}) is known to be chaotic \citep{Avishai2002,Santos2004:Entangelement,Santos2004i:Integrability}.

To study the classical limit, it is convenient to normalize the spin
operators, $\hat{\mathcal{S}}_{\alpha}^{k}=\hat{S}_{\alpha}^{k}/\sqrt{s\left(s+1\right)}$,
which amounts to fixing the largest eigenvalue of the $\hat{\mathcal{S}}^{2}=\sum_{k}\left(\hat{\mathcal{S}}_{\alpha}^{k}\right)^{2}$
operator to 1. Using the normalized spin operators, the quantum Hamiltonian
is given by 
\begin{align}
\hat{H} & =J_{xy}\sum_{k=1}^{L-1}\left(\hat{\mathcal{S}}_{x}^{k}\hat{\mathcal{S}}_{x}^{k+1}+\hat{\mathcal{S}}_{y}^{k}\hat{\mathcal{S}}_{y}^{k+1}\right)\\
 & +J_{z}\sum_{k=1}^{L-1}\hat{\mathcal{S}}_{z}^{k}\hat{\mathcal{S}}_{z}^{k+1}+\sum_{i=1}^{L}h_{k}\hat{\mathcal{S}}_{z}^{k},\nonumber 
\end{align}
where the commutation relations of the \emph{normalized} spins follow
directly from the standard commutation relations of the spin,
\begin{equation}
\left[\hat{\mathcal{S}}_{\alpha}^{k},\hat{\mathcal{S}}_{\beta}^{k}\right]=i\frac{1}{\sqrt{s\left(s+1\right)}}\epsilon_{\alpha\beta\gamma}\hat{\mathcal{S}}_{\gamma}^{k},
\end{equation}
where $\epsilon_{\alpha\beta\gamma}$ is the Levi-Civita symbol. In
the $s\to\infty$ limit the normalized spins commute, which corresponds
to the classical limit. We can then replace the operators $\hat{\mathcal{S}}_{\alpha}^{k}$
by real numbers $s_{\alpha}^{k}$, and obtain the classical version
of the disordered XXZ model, 
\begin{align}
H_{cl} & =J_{xy}\sum_{k=1}^{L-1}\left(s_{x}^{k}s_{x}^{k+1}+s_{y}^{k}s_{y}^{k+1}\right)\label{eq:XXZ classical limit}\\
 & +J_{z}\sum_{k=1}^{L-1}s_{z}^{k}s_{z}^{k+1}+\sum_{i=1}^{L}h_{k}s_{z}^{k},\nonumber 
\end{align}
which represents classical interacting rotators $\vec{s}_{i}=\left(s_{x}^{i},s_{y}^{i},s_{z}^{i}\right)$
on a unit sphere. The classical system also conserves the total magnetization.
We start by studying the chaotic properties of the model in the classical
limit.

\section{Classical chaos}

To examine the chaotic properties of the classical Hamiltonian $H_{cl}$
in Eq.~(\ref{eq:XXZ classical limit}) , we examine the Lyapunov
exponents starting from all the rotators pointing down in the $z$-direction,
which corresponds to the lowest magnetization limit. The equations
of motion of the rotators are obtained using Eq.~(\ref{eq:XXZ classical limit})
and Poisson brackets, 
\begin{equation}
\frac{ds_{\alpha}^{k}}{dt}=\left\{ H_{cl},s_{\alpha}^{k}\right\} ,\qquad\left\{ s_{\alpha}^{k},s_{\beta}^{k}\right\} =\delta_{kl}\epsilon_{\alpha\beta\gamma}s_{\gamma}^{k},
\end{equation}
which gives
\begin{align}
\frac{ds_{x}^{k}}{dt} & =-J_{xy}\left(s_{y}^{k+1}+s_{y}^{k-1}\right)s_{z}^{k}+\left[J_{z}\left(s_{z}^{k+1}+s_{z}^{k-1}\right)+h_{k}\right]s_{y}^{k}\nonumber \\
\frac{ds_{y}^{k}}{dt} & =J_{xy}\left(s_{x}^{k-1}+s_{x}^{k+1}\right)s_{z}^{k}-\left[J_{z}\left(s_{z}^{k+1}+s_{z}^{k-1}\right)+h_{k}\right]s_{x}^{k}\nonumber \\
\frac{ds_{z}^{k}}{dt} & =-J_{xy}\left[\left(s_{x}^{k-1}+s_{x}^{k+1}\right)s_{y}^{k}-\left(s_{y}^{k-1}+s_{y}^{k+1}\right)s_{x}^{k}\right],
\end{align}
and can be compactly written as nonlinear Bloch equations,
\begin{equation}
\frac{d\vec{s}_{k}}{dt}=-\vec{b}_{k}\times\vec{s}_{k},
\end{equation}
with an effective magnetic field,
\begin{equation}
\vec{b}_{k}=\vec{h}_{k}+\vec{J}\cdot\left(\vec{s}_{k+1}+\vec{s}_{k-1}\right),
\end{equation}
where we defined the coupling vector $\vec{J}=\left(J_{xy},J_{xy},J_{z}\right)$.
Since the equations conserve $s_{k}^{2}=\vec{s}_{k}\cdot\vec{s}_{k}=1$
of each rotator separately, the equation of one of the components
of $\vec{s}$ is redundant and it is advantageous to use a numerical
integration scheme which conserves all $s_{k}^{2}$ explicitly. One
way to do this is to parametrize the orientation of the rotators on
the unit sphere as $\vec{s}_{k}=\left(\sin\theta_{k}\cos\phi_{k},\sin\theta_{k}\sin\phi_{k},\cos\theta_{k}\right)$.
This yields the following equations of motion for the angles,
\begin{align}
\frac{d\phi_{k}}{dt} & =J_{xy}\sin\theta_{k-1}\cot\theta_{k}\left(\cos\phi_{k-1}\cos\phi_{k}+\sin\phi_{k-1}\sin\phi_{k}\right)\nonumber \\
 & +J_{xy}\sin\theta_{k+1}\cot\theta_{k}\left(\cos\phi_{k+1}\cos\phi_{k}+\sin\phi_{k+1}\sin\phi_{k}\right)\nonumber \\
 & -J_{z}\left(\cos\theta_{k+1}+\cos\theta_{k-1}\right)-h_{k},\label{eq:classical equations of motion}\\
\frac{d\theta_{k}}{dt} & =J_{xy}\sin\theta_{k+1}\left(\cos\phi_{k+1}\sin\phi_{k}-\sin\phi_{k+1}\cos\phi_{k}\right)\nonumber \\
 & +J_{xy}\sin\theta_{k-1}\left(\cos\phi_{k-1}\sin\phi_{k}-\sin\phi_{k-1}\cos\phi_{k}\right).\nonumber 
\end{align}
To calculate the Lyapunov exponents of this system we initialize all
of the rotors with a slight deflection from the $z$ axis, $\delta\theta=\pi-\theta$,
which corresponds to the low magnetization setting considered in this
work. We set the angles $\phi_{i}$ in the $xy$ plane to be randomized,
and integrate the equations of motion. The maximal Lyapunov exponent
is calculated with the algorithm proposed in Ref.~\citep{Elsayed2014},
which is based on finding the rate of growth of the fastest growing
tangent space vector. The initial conditions in the azimuthal direction
and the on-site disorder are changed randomly between realizations
and the maximal exponents are then averaged. We use a chain of 50
spins to calculate the Lyapunov exponents and integrate the dynamics
using the LSODA method described in Ref.~\citep{Petzold1980} with
time steps chosen optimally by the algorithm.

\begin{figure}[th]
\begin{centering}
\includegraphics{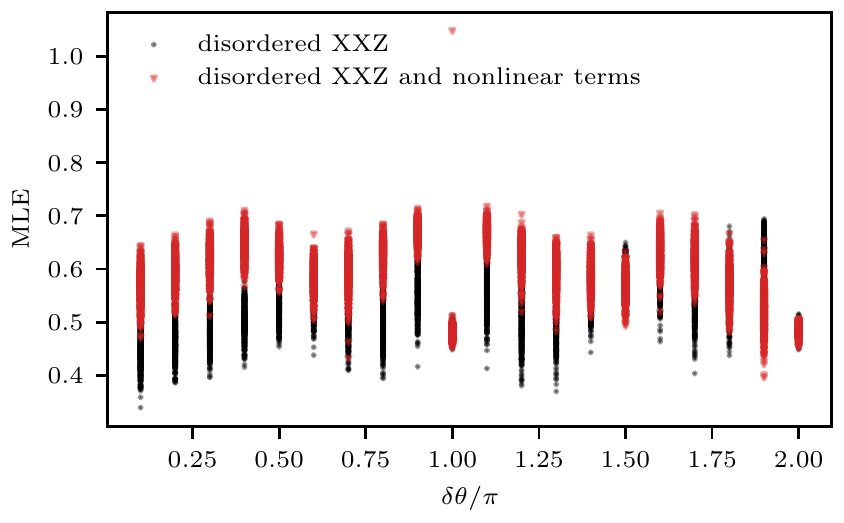}
\par\end{centering}
\caption{Maximal Lyapunov exponent (MLE) as a function of the deviation $\delta\theta$
(measured in radians) from the initial condition with all the spins
pointing in the $z$-direction. Each point represents the average
over different realizations for random values of $\phi_{i}$ and for
the on-site disorder. The black circles correspond to the maximal
Lyapunov exponent of the model in Eq.~(\ref{eq:XXZ classical limit}),
while the red triangles correspond to the maximal Lyapunov exponent
of the model in Eq.~(\ref{eq:classical-with-nonlinear-terms}) with
additional nonlinear terms. \label{fig:mles} A chain of $L=50$ is
considered for these simulations.}
\end{figure}

In Fig.~\ref{fig:mles} we present the results of the calculation
for different deviations $\delta\theta$. It can be seen that even
for low values of $\delta\theta$, which is equivalent to a low number
of excitations in the quantum limit, the maximal Lyapunov exponent
is positive, indicating that the classical limit is chaotic.

\section{Quantum chaos}

After establishing that the classical limit of our model in Eq.~(\ref{eq:XXZ classical limit})
is chaotic at low magnetization, we proceed to examine whether the
quantum Hamiltonian in Eq.~(\ref{eq:XXZ model}) exhibits properties
associated with quantum chaos. These properties include correlated
eigenvalues, as in random matrix theory, and eigenstates that away
from the edges of the spectrum are close to the eigenstates of full
random matrices, that is, their components are nearly independent
real random numbers from a Gaussian distribution satisfying the normalization
condition \citep{Zelevinsky1996,Borgonovi2016}. The onset of these
almost random vectors in many-body quantum systems results in normal
distributions of the off-diagonal elements of local observables \citep{Pechukas1983,Pechukas1984,Peres1984,Peres1984a,Feingold1984,Feingold1985,Feingold1986},
which is one of the features of the eigenstate thermalization hypothesis
(ETH) \citep{Deutsch1991,Srednicki1994,Srednicki1995,Srednicki1999,Rigol2008,Beugeling2015,Leblond2019,Khaymovich2019,Santos2020,ydba2021}.

Our analysis of quantum chaos focuses on level statistics and the
off-diagonal ETH. It is done for the subspace with total $z$-magnetization
equal to $-sL+N$, where $N$ is a fixed number of excitations.

To study the level statistics, we use the so called $r$-metric~\citep{Oganesyan2007,Atas2013,Corps2020},
\begin{equation}
r_{\alpha}=\min\left(\frac{\delta_{\alpha}}{\delta_{\alpha-1}},\frac{\delta_{\alpha-1}}{\delta_{\alpha}}\right),\label{eq:rn def}
\end{equation}
where $\delta_{\alpha}=E_{\alpha+1}-E_{\alpha}$ is the spacing between
the neighboring eigenvalues of the Hamiltonian. The $r$-metric captures
short-range correlations among the energy levels. For an integrable
system with Poissonian level spacing distribution, the average of
$r_{\alpha}$ over the spectrum gives $r_{\text{Poisson}}\approx0.39$,
while for chaotic systems $r_{\text{GOE}}\approx0.53$ \citep{Atas2013}
The latter is the value obtained for full random matrices from Gaussian
orthogonal ensembles (GOE) \citep{Atas2013}.
\begin{figure}[t]
\begin{centering}
\includegraphics{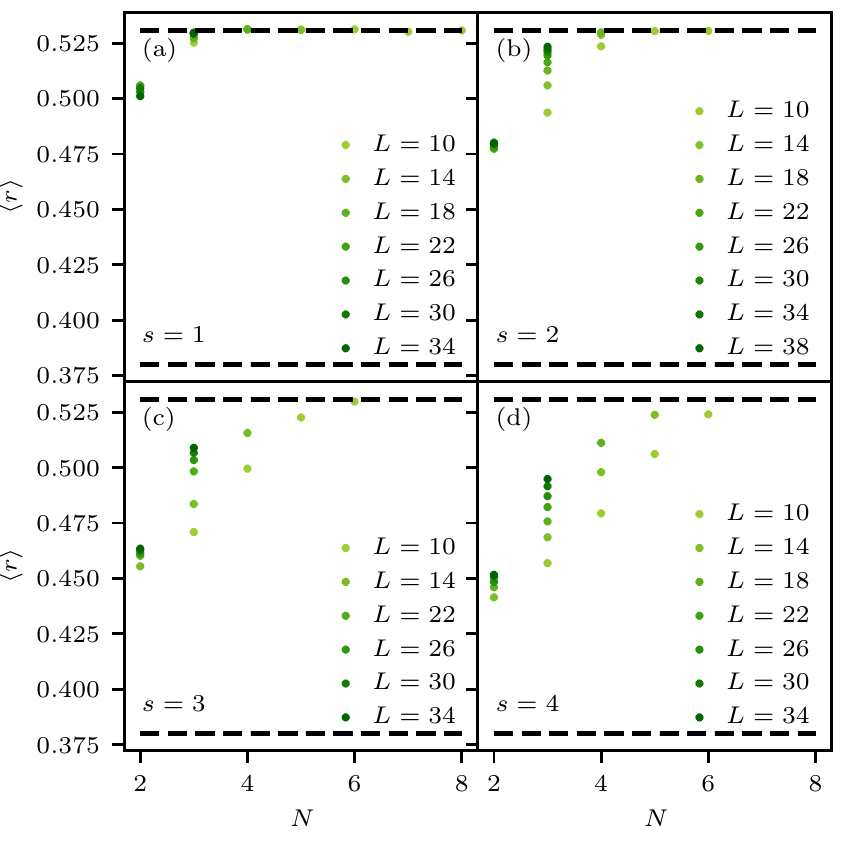}
\par\end{centering}
\caption{Average of the $r$-metric for four different spin sizes, $s=1,2,3,4$,
as a function of the number of excitations $N$ and various system
sizes. Darker circles represent larger chain sizes (see legends).
In the calculations of this metric, the top and bottom 15\% of the
eigenvalues are omitted. The dashed horizontal lines stand for $r_{\text{Poisson}}=0.39$
and $r_{\text{GOE}}=0.53$. \label{fig:rstats_disorder}}
\end{figure}

In Fig.~\ref{fig:rstats_disorder} we plot the $r$-metric for four
different spin sizes as a function of the number $N$ of excitations
in the system. We repeat our calculations for chain lengths ranging
from 10 to 38 sites (darker colors indicate larger chains in the figure).
Surprisingly, while the classical limit of this model is chaotic,
as shown in the previous section, for larger spin sizes, \emph{more}
excitations are required to reach signatures of chaos in the quantum
domain. The system with spin 1 is fairly chaotic for $N=3$ and 4,
similarly to the spin-1/2 model studied in Refs.~\citep{schiulaz2018:FewManybodyQuantum,Zisling2021},
but $\left\langle r\right\rangle $ for the spin-4 model does not
reach $r_{\text{GOE}}$ for the system sizes considered. For a fixed
system size and a fixed number of excitations, the $r$-metric in
Fig.~\ref{fig:rstats_disorder} indicates that a system with a larger
spin presents a weaker degree of chaos than its counterpart with a
smaller spin.

\begin{figure}[th]
\begin{centering}
\includegraphics{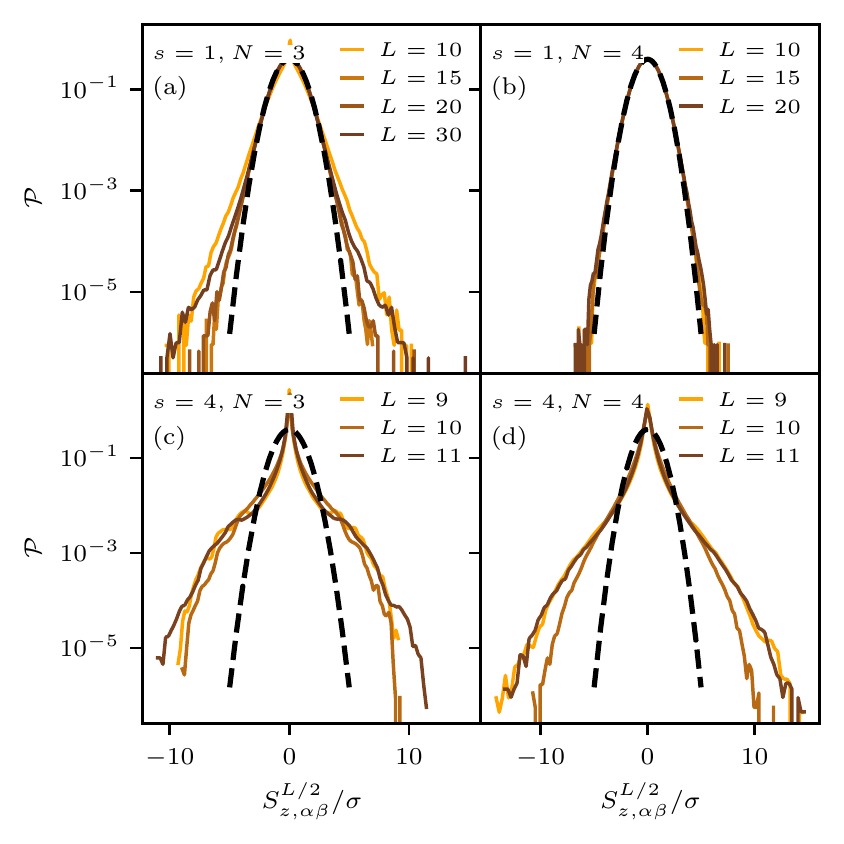}
\par\end{centering}
\caption{Distributions of the off-diagonal elements of the $\hat{S}_{z}^{L/2}$
operator for spin 1 (top row) and spin 4 (bottom row) and varying
system sizes. Left column corresponds to $N=3$ excitations and right
column to $N=4$ excitations. We use 250 eigenstates in the middle
of the spectrum to compute the off-diagonal elements, and normalize
the distributions to have unit variance. The dashed black line shows
a Gaussian distribution with unit variance.\label{fig:distribution_disorder}}
\end{figure}

To better understand why the degree of quantum chaos decreases as
the spin is enlarged, we resort to the analysis of the distribution
of the off-diagonal elements of the local magnetization, $\hat{S}_{z}^{L/2}$.
For this purpose we take 250 states from the middle of the spectrum
and compute the matrix element $S_{z,\alpha\beta}^{L/2}\equiv\bra{\alpha}\hat{S}_{z}^{L/2}\ket{\beta}$
where $\alpha\neq\beta$ are two different eigenstates. Since the
variance of the distribution decreases with the Hilbert space dimension
$\mathcal{D}$ \citep{Beugeling2015}, we normalize the distribution
by dividing $S_{z,\alpha\beta}^{L/2}$ by its standard deviation,
$\sigma=\text{std}\left(S_{z,\alpha\beta}^{L/2}\right)$.

\begin{figure}[t]
\begin{centering}
\includegraphics{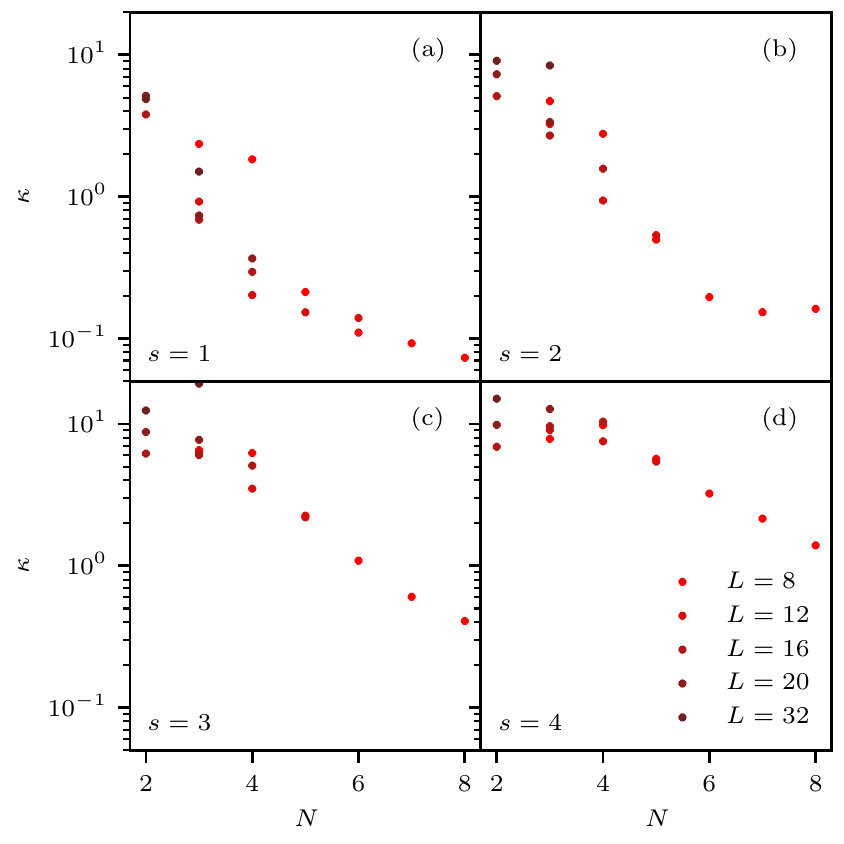}
\par\end{centering}
\caption{Kurtosis of the distributions of the off-diagonal elements of the
$\hat{S}_{z}^{L/2}$ operator for four different spin sizes, $s=1,2,3,4$,
as a function of the number of excitations $N$ and various system
sizes. Darker circles represent larger chain sizes (see legends).
In the calculations of the kurtosis we used 250 eigenstates from the
middle of the spectrum. For a chaotic system the kurtosis is supposed
to be zero. \label{fig:kurtosis_disorder}}
\end{figure}

The resulting distributions for $s=1$, $4$ and $N=3$,$4$ can be
seen in Fig.~\ref{fig:distribution_disorder}. The distributions
for the spin-1 case in Figs.~\ref{fig:distribution_disorder}~(a)-(b)
are very close to Gaussian (dashed black line) for both $N=3$ {[}Fig.~\ref{fig:distribution_disorder}~(a){]}
and $N=4$ {[}Fig.~\ref{fig:distribution_disorder}~(b){]} excitations.
However, the spin-4 distributions depicted in in Fig.~\ref{fig:distribution_disorder}~(c)-(d)
are strongly peaked around zero. To quantify how close the distributions
are to a Gaussian, we calculate their normalized kurtosis,
\begin{equation}
\text{\ensuremath{\kappa}}=\frac{\langle O_{\alpha\beta}^{4}\rangle-\langle O_{\alpha\beta}\rangle^{4}}{\sigma^{4}}-3,\label{eq:kurtosis def}
\end{equation}
which is equal to 0 for a Gaussian distribution. Figure~\ref{fig:kurtosis_disorder}
shows the kurtosis of the distributions for $s=1,2,3$ and $4$ as
a function of the number of excitations $N$ for different system
sizes. For $s=3$ and $4$ the dependence of the kurtosis on the number
of excitations is non-monotonic around $N=2,3$. For $s=4$, the kurtosis,
$\kappa$, only decreases when $N>4$, and much slower than for $s=1,2$.
These results confirm that larger spins require more excitations to
reach the chaotic regime. Note that the situation does not improve
with system size, since for larger $L$, the kurtosis actually moves
further away from zero.

Why do larger spins result in a reduced level of quantum chaos? Large-spin
chains can contain sites with many excitations, while our disordered
XXZ model in Eq.~(\ref{eq:XXZ model}) allows only for the motion
of one excitation at a time, which makes these clusters of excitations
very hard to disassemble. We conjecture that these large and slow
moving clusters prevent the onset of strong chaos in our system and
propose two mechanisms to reverse this scenario, as discussed next.

\section{Enhancing quantum chaos}

In this section we consider two ways to improve the chaotic properties
of our finite large-spin system by modifying its Hamiltonian. The
first mechanism that we consider is the diversification of the sizes
of the clusters of excitations by giving different energies to clusters
with different number of excitations. This is achieved with the addition
of properly normalized nonlinear magnetization terms to the Hamiltonian,

\begin{align}
H_{1} & =H+\frac{\alpha}{s\left(s+1\right)}\sum_{k}\left(\hat{S}_{z}^{k}\right)^{2},\label{eq:H1}\\
 & +\frac{\mu}{\left(s\left(s+1\right)\right)^{3/2}}\sum_{k}\left(\hat{S}_{z}^{k}\right)^{3},\nonumber 
\end{align}
where $\alpha=0.87,\mu=0.91$, and similarly to Eq.~(\ref{eq:XXZ model}),
the normalization of the nonlinear terms is taken to pertain a proper
classical limit for $s\to\infty$. The classical limit of this model
is
\begin{equation}
H_{1}^{cl}=H_{cl}+\alpha\sum_{k}\left(s_{z}^{k}\right)^{2}+\mu\sum_{k}\left(s_{z}^{k}\right)^{3}.\label{eq:classical-with-nonlinear-terms}
\end{equation}
 In Fig.~\ref{fig:mles}, we present the maximal Lyapunov exponents
of this model with red points and verify that they closely follow
the Lyapunov exponents of the classical disordered XXZ model in Eq.~(\ref{eq:XXZ classical limit}).
We note that the addition of these terms slightly increases the Lyapunov
exponents.

The second quantum chaos enhancement mechanism that we consider is
the facilitation of the fragmentation of the clusters of excitations
using nonlinear ladder operators, 
\begin{align}
H_{2} & =H_{1}+\label{eq:H2}\\
 & +\sum_{k=1}^{L-1}\sum_{n=2}^{s}\frac{J_{xy}}{s^{n}\left(s+1\right)^{n}}\left[\left(\hat{S}_{+}^{k}\hat{S}_{-}^{k+1}\right)^{n}+\left(\hat{S}_{+}^{k+1}\hat{S}_{-}^{k}\right)^{n}\right].\nonumber 
\end{align}
The added terms move $k-$excitations between neighboring sites with
$2\leq k\leq s$. While the classical limit of this model is well
defined, the derivation of its Hamiltonian in a closed form is cumbersome,
because the sum over $k$ of the nonlinear ladder operators has to
be computed explicitly, so we do not show it here.

\begin{figure}[t]
\begin{centering}
\includegraphics{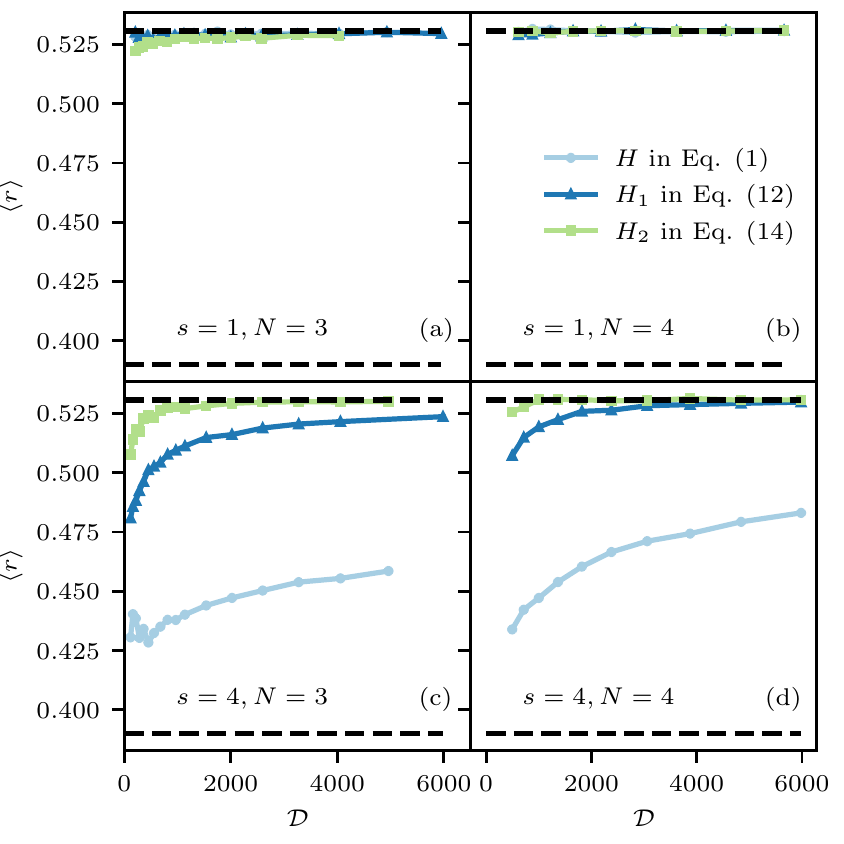}
\par\end{centering}
\caption{Averaged $r$-metric as a function of the Hilbert space dimension
$\mathcal{D}$ for spin-1 models(top row) and spin-4 models (bottom
row). The left column shows $N=3$ excitations and the right column
shows $N=4$ excitations. Circles correspond to the disordered XXZ
model (\ref{eq:XXZ model}), triangles to model (\ref{eq:H1}), and
squares to model (\ref{eq:H2}). The dashed horizontal lines stand
for $r_{\text{Poisson}}=0.39$ and $r_{\text{GOE}}=0.5307$. \label{fig:added_terms}}
\end{figure}

The results for the Hamiltonians in Eq.~(\ref{eq:H1}) and Eq.~(\ref{eq:H2})
show a systematic improvement for all considered quantum chaos metrics
when compared to the disordered XXZ model in Eq.~(\ref{eq:XXZ model}).
Figure~\ref{fig:added_terms} shows the $r$-metric of all three
models computed for spin 1 {[}Figs.~\ref{fig:added_terms}(a)-(b){]}
and spin 4 Figs.~\ref{fig:added_terms}(c)-(d){]} as a function of
the Hilbert space dimension, $\mathcal{D}$. The spin-1 case is already
fairly chaotic for the Hamiltonian (\ref{eq:XXZ model}), so the addition
of the new terms in Eq.~(\ref{eq:H1}) and Eq.~(\ref{eq:H2}) do
not affect the values of $\left\langle r\right\rangle $. However,
for spin 4, adding the nonlinear magnetization terms in Eq.~(\ref{eq:H1})
dramatically improves the degree of quantum chaos for both $N=3$
and 4. The addition of the nonlinear ladder operators in Eq.~(\ref{eq:H2})
is even more effective and leads to strong level repulsion for as
few as 3 excitations, making the results analogous to the case of
spin-1/2 studied in Ref.~\citep{Zisling2021}. This significant enhancement
of the degree of quantum chaos is corroborated by the other chaotic
metrics considered in this work, namely the distributions of off-diagonal
observables and the kurtosis of these distributions (not shown).

\section{Discussion}

The driving question of this work is whether increasing the spin size
of a one-dimensional spin model can reduce the number of spin excitations
needed to achieve quantum chaos. For this purpose we studied the large
spin limit of a disordered XXZ chain. For a single spin-1/2 excitation
this model is integrable and localized via the Anderson localization
mechanism, while at zero $z$-magnetization it is chaotic. In previous
studies of spin-1/2 chains, it was established that at least 3 spin
excitations are required to achieve quantum chaos. By considering
chaotic metrics based on both the eigenvalues and the eigenstates,
we found that, although the classical limit of the disordered XXZ
chain is chaotic for very low magnetization, the large spin version
of the quantum XXZ model, surprisingly, shows significantly reduced
chaotic behavior, compared to its spin-1/2 counterpart. We attribute
this phenomenon to the occurrence of clusters of excitations concentrated
on one or a number of neighboring sites, which makes the clusters
hard to decompose and the corresponding dynamics slow. We proposed
two mechanisms to enhance quantum chaotic behavior: by the introduction
of additional relaxation channels for the clusters in the face of
a collection of ladder operators and by adding nonlinear onsite magnetization.
We numerically verified that the introduction of these terms allows
to achieve quantum chaos for $N=3$ excitations, similarly to the
spin-1/2 systems. Interestingly, the introduction of additional relaxation
channels enhances \emph{classical} chaos only slightly. Our study
suggests that care should be taken in considering large-spin limits
of quantum models, since certain relaxation channels become inherently
slow. It also suggests that classical chaos and quantum chaos might
not be so tightly bound.
\selectlanguage{american}%
\begin{acknowledgments}
This research was supported by a grant from the United States-Israel
Binational Foundation (BSF, Grant No. $2019644$), Jerusalem, Israel,
and the United States National Science Foundation (NSF, Grant No.
DMR$-1936006$), and by the Israel Science Foundation (grants No.
527/19 and 218/19).
\end{acknowledgments}

\selectlanguage{english}%
\bibliography{refs/lib_ybl,refs/lib_lea,refs/lib_yael}

\begin{thebibliography}{54}%
\makeatletter
\providecommand \@ifxundefined [1]{%
 \@ifx{#1\undefined}
}%
\providecommand \@ifnum [1]{%
 \ifnum #1\expandafter \@firstoftwo
 \else \expandafter \@secondoftwo
 \fi
}%
\providecommand \@ifx [1]{%
 \ifx #1\expandafter \@firstoftwo
 \else \expandafter \@secondoftwo
 \fi
}%
\providecommand \natexlab [1]{#1}%
\providecommand \enquote  [1]{``#1''}%
\providecommand \bibnamefont  [1]{#1}%
\providecommand \bibfnamefont [1]{#1}%
\providecommand \citenamefont [1]{#1}%
\providecommand \href@noop [0]{\@secondoftwo}%
\providecommand \href [0]{\begingroup \@sanitize@url \@href}%
\providecommand \@href[1]{\@@startlink{#1}\@@href}%
\providecommand \@@href[1]{\endgroup#1\@@endlink}%
\providecommand \@sanitize@url [0]{\catcode `\\12\catcode `\$12\catcode
  `\&12\catcode `\#12\catcode `\^12\catcode `\_12\catcode `\%12\relax}%
\providecommand \@@startlink[1]{}%
\providecommand \@@endlink[0]{}%
\providecommand \url  [0]{\begingroup\@sanitize@url \@url }%
\providecommand \@url [1]{\endgroup\@href {#1}{\urlprefix }}%
\providecommand \urlprefix  [0]{URL }%
\providecommand \Eprint [0]{\href }%
\providecommand \doibase [0]{https://doi.org/}%
\providecommand \selectlanguage [0]{\@gobble}%
\providecommand \bibinfo  [0]{\@secondoftwo}%
\providecommand \bibfield  [0]{\@secondoftwo}%
\providecommand \translation [1]{[#1]}%
\providecommand \BibitemOpen [0]{}%
\providecommand \bibitemStop [0]{}%
\providecommand \bibitemNoStop [0]{.\EOS\space}%
\providecommand \EOS [0]{\spacefactor3000\relax}%
\providecommand \BibitemShut  [1]{\csname bibitem#1\endcsname}%
\let\auto@bib@innerbib\@empty
%</preamble>
\bibitem [{\citenamefont {Casati}\ \emph {et~al.}(1980)\citenamefont {Casati},
  \citenamefont {Valz-Gris},\ and\ \citenamefont {Guarnieri}}]{Casati1980}%
  \BibitemOpen
  \bibfield  {author} {\bibinfo {author} {\bibfnamefont {G.}~\bibnamefont
  {Casati}}, \bibinfo {author} {\bibfnamefont {F.}~\bibnamefont {Valz-Gris}},\
  and\ \bibinfo {author} {\bibfnamefont {I.}~\bibnamefont {Guarnieri}},\
  }\bibfield  {title} {\bibinfo {title} {On the connection between quantization
  of nonintegrable systems and statistical theory of spectra},\ }\href
  {https://doi.org/10.1007/bf02798790} {\bibfield  {journal} {\bibinfo
  {journal} {Lett. Nuovo Cimento}\ }\textbf {\bibinfo {volume} {28}},\ \bibinfo
  {pages} {279} (\bibinfo {year} {1980})}\BibitemShut {NoStop}%
\bibitem [{\citenamefont {Bohigas}\ \emph {et~al.}(1984)\citenamefont
  {Bohigas}, \citenamefont {Giannoni},\ and\ \citenamefont
  {Schmit}}]{Bohigas1984}%
  \BibitemOpen
  \bibfield  {author} {\bibinfo {author} {\bibfnamefont {O.}~\bibnamefont
  {Bohigas}}, \bibinfo {author} {\bibfnamefont {M.~J.}\ \bibnamefont
  {Giannoni}},\ and\ \bibinfo {author} {\bibfnamefont {C.}~\bibnamefont
  {Schmit}},\ }\bibfield  {title} {\bibinfo {title} {Characterization of
  chaotic quantum spectra and universality of level fluctuation laws},\ }\href
  {https://doi.org/10.1103/physrevlett.52.1} {\bibfield  {journal} {\bibinfo
  {journal} {Phys. Rev. Lett.}\ }\textbf {\bibinfo {volume} {52}},\ \bibinfo
  {pages} {1} (\bibinfo {year} {1984})}\BibitemShut {NoStop}%
\bibitem [{\citenamefont {Stockmann}(2007)}]{Stockmann2007}%
  \BibitemOpen
  \bibfield  {author} {\bibinfo {author} {\bibfnamefont {H.-J.}\ \bibnamefont
  {Stockmann}},\ }\href@noop {} {\emph {\bibinfo {title} {Quantum Chaos}}}\
  (\bibinfo  {publisher} {Cambridge University Press},\ \bibinfo {address}
  {Cambridge, England},\ \bibinfo {year} {2007})\BibitemShut {NoStop}%
\bibitem [{\citenamefont {Haake}(2001)}]{Haake2001}%
  \BibitemOpen
  \bibfield  {author} {\bibinfo {author} {\bibfnamefont {F.}~\bibnamefont
  {Haake}},\ }\href@noop {} {\emph {\bibinfo {title} {Quantum Signatures of
  Chaos}}},\ \bibinfo {edition} {2nd}\ ed.,\ Springer Series in Synergetics\
  (\bibinfo  {publisher} {Springer},\ \bibinfo {address} {Berlin, Germany},\
  \bibinfo {year} {2001})\BibitemShut {NoStop}%
\bibitem [{\citenamefont {Reichl}(1994)}]{Reichl1994}%
  \BibitemOpen
  \bibfield  {author} {\bibinfo {author} {\bibfnamefont {L.~E.}\ \bibnamefont
  {Reichl}},\ }\href@noop {} {\emph {\bibinfo {title} {The transition to
  chaos}}},\ Institute for Nonlinear Science\ (\bibinfo  {publisher}
  {Springer},\ \bibinfo {address} {New York, NY},\ \bibinfo {year}
  {1994})\BibitemShut {NoStop}%
\bibitem [{\citenamefont {Zelevinsky}\ \emph {et~al.}(1996)\citenamefont
  {Zelevinsky}, \citenamefont {Brown}, \citenamefont {Frazier},\ and\
  \citenamefont {Horoi}}]{Zelevinsky1996}%
  \BibitemOpen
  \bibfield  {author} {\bibinfo {author} {\bibfnamefont {V.}~\bibnamefont
  {Zelevinsky}}, \bibinfo {author} {\bibfnamefont {B.}~\bibnamefont {Brown}},
  \bibinfo {author} {\bibfnamefont {N.}~\bibnamefont {Frazier}},\ and\ \bibinfo
  {author} {\bibfnamefont {M.}~\bibnamefont {Horoi}},\ }\bibfield  {title}
  {\bibinfo {title} {The nuclear shell model as a testing ground for many-body
  quantum chaos},\ }\href {https://doi.org/10.1016/s0370-1573(96)00007-5}
  {\bibfield  {journal} {\bibinfo  {journal} {Phys. Rep.}\ }\textbf {\bibinfo
  {volume} {276}},\ \bibinfo {pages} {85} (\bibinfo {year} {1996})}\BibitemShut
  {NoStop}%
\bibitem [{\citenamefont {Borgonovi}\ \emph {et~al.}(2016)\citenamefont
  {Borgonovi}, \citenamefont {Izrailev}, \citenamefont {Santos},\ and\
  \citenamefont {Zelevinsky}}]{Borgonovi2016}%
  \BibitemOpen
  \bibfield  {author} {\bibinfo {author} {\bibfnamefont {F.}~\bibnamefont
  {Borgonovi}}, \bibinfo {author} {\bibfnamefont {F.~M.}\ \bibnamefont
  {Izrailev}}, \bibinfo {author} {\bibfnamefont {L.~F.}\ \bibnamefont
  {Santos}},\ and\ \bibinfo {author} {\bibfnamefont {V.~G.}\ \bibnamefont
  {Zelevinsky}},\ }\bibfield  {title} {\bibinfo {title} {Quantum chaos and
  thermalization in isolated systems of interacting particles},\ }\href
  {https://doi.org/10.1016/j.physrep.2016.02.005} {\bibfield  {journal}
  {\bibinfo  {journal} {Phys. Rep.}\ }\textbf {\bibinfo {volume} {626}},\
  \bibinfo {pages} {1} (\bibinfo {year} {2016})}\BibitemShut {NoStop}%
\bibitem [{\citenamefont {D'Alessio}\ \emph {et~al.}(2016)\citenamefont
  {D'Alessio}, \citenamefont {Kafri}, \citenamefont {Polkovnikov},\ and\
  \citenamefont {Rigol}}]{Alessio2016}%
  \BibitemOpen
  \bibfield  {author} {\bibinfo {author} {\bibfnamefont {L.}~\bibnamefont
  {D'Alessio}}, \bibinfo {author} {\bibfnamefont {Y.}~\bibnamefont {Kafri}},
  \bibinfo {author} {\bibfnamefont {A.}~\bibnamefont {Polkovnikov}},\ and\
  \bibinfo {author} {\bibfnamefont {M.}~\bibnamefont {Rigol}},\ }\bibfield
  {title} {\bibinfo {title} {From quantum chaos and eigenstate thermalization
  to statistical mechanics and thermodynamics},\ }\href
  {https://doi.org/10.1080/00018732.2016.1198134} {\bibfield  {journal}
  {\bibinfo  {journal} {Adv. Phys.}\ }\textbf {\bibinfo {volume} {65}},\
  \bibinfo {pages} {239} (\bibinfo {year} {2016})}\BibitemShut {NoStop}%
\bibitem [{\citenamefont {D'Alessio}\ \emph {et~al.}(2014)\citenamefont
  {D'Alessio}, \citenamefont {Rigol}, \citenamefont {D'Alessio},\ and\
  \citenamefont {Rigol}}]{DAlessio2014}%
  \BibitemOpen
  \bibfield  {author} {\bibinfo {author} {\bibfnamefont {L.}~\bibnamefont
  {D'Alessio}}, \bibinfo {author} {\bibfnamefont {M.}~\bibnamefont {Rigol}},
  \bibinfo {author} {\bibfnamefont {L.}~\bibnamefont {D'Alessio}},\ and\
  \bibinfo {author} {\bibfnamefont {M.}~\bibnamefont {Rigol}},\ }\bibfield
  {title} {\bibinfo {title} {Long-time {{Behavior}} of {{Isolated Periodically
  Driven Interacting Lattice Systems}}},\ }\href
  {https://doi.org/10.1103/PhysRevX.4.041048} {\bibfield  {journal} {\bibinfo
  {journal} {Phys. Rev. X}\ }\textbf {\bibinfo {volume} {4}},\ \bibinfo {pages}
  {041048} (\bibinfo {year} {2014})}\BibitemShut {NoStop}%
\bibitem [{\citenamefont {Lazarides}\ \emph {et~al.}(2015)\citenamefont
  {Lazarides}, \citenamefont {Das},\ and\ \citenamefont
  {Moessner}}]{Lazarides2014}%
  \BibitemOpen
  \bibfield  {author} {\bibinfo {author} {\bibfnamefont {A.}~\bibnamefont
  {Lazarides}}, \bibinfo {author} {\bibfnamefont {A.}~\bibnamefont {Das}},\
  and\ \bibinfo {author} {\bibfnamefont {R.}~\bibnamefont {Moessner}},\
  }\bibfield  {title} {\bibinfo {title} {Fate of {{Many}}-{{Body Localization
  Under Periodic Driving}}},\ }\href
  {https://doi.org/10.1103/PhysRevLett.115.030402} {\bibfield  {journal}
  {\bibinfo  {journal} {Phys. Rev. Lett.}\ }\textbf {\bibinfo {volume} {115}},\
  \bibinfo {pages} {030402} (\bibinfo {year} {2015})}\BibitemShut {NoStop}%
\bibitem [{\citenamefont {Nandkishore}\ and\ \citenamefont
  {Huse}(2015)}]{Nandkishore2014}%
  \BibitemOpen
  \bibfield  {author} {\bibinfo {author} {\bibfnamefont {R.}~\bibnamefont
  {Nandkishore}}\ and\ \bibinfo {author} {\bibfnamefont {D.~A.}\ \bibnamefont
  {Huse}},\ }\bibfield  {title} {\bibinfo {title} {Many-{{Body Localization}}
  and {{Thermalization}} in {{Quantum Statistical Mechanics}}},\ }\href
  {https://doi.org/10.1146/annurev-conmatphys-031214-014726} {\bibfield
  {journal} {\bibinfo  {journal} {Annu. Rev. Condens. Matter Phys.}\ }\textbf
  {\bibinfo {volume} {6}},\ \bibinfo {pages} {15} (\bibinfo {year}
  {2015})}\BibitemShut {NoStop}%
\bibitem [{\citenamefont {Abanin}\ and\ \citenamefont
  {Papi{\'c}}(2017)}]{Abanin2017}%
  \BibitemOpen
  \bibfield  {author} {\bibinfo {author} {\bibfnamefont {D.~A.}\ \bibnamefont
  {Abanin}}\ and\ \bibinfo {author} {\bibfnamefont {Z.}~\bibnamefont
  {Papi{\'c}}},\ }\bibfield  {title} {\bibinfo {title} {Recent progress in
  many-body localization},\ }\href {https://doi.org/10.1002/andp.201700169}
  {\bibfield  {journal} {\bibinfo  {journal} {Ann. Phys.}\ }\textbf {\bibinfo
  {volume} {529}},\ \bibinfo {pages} {1700169} (\bibinfo {year}
  {2017})}\BibitemShut {NoStop}%
\bibitem [{\citenamefont {Agarwal}\ \emph {et~al.}(2017)\citenamefont
  {Agarwal}, \citenamefont {Altman}, \citenamefont {Demler}, \citenamefont
  {Gopalakrishnan}, \citenamefont {Huse},\ and\ \citenamefont
  {Knap}}]{Agarwal2016_review}%
  \BibitemOpen
  \bibfield  {author} {\bibinfo {author} {\bibfnamefont {K.}~\bibnamefont
  {Agarwal}}, \bibinfo {author} {\bibfnamefont {E.}~\bibnamefont {Altman}},
  \bibinfo {author} {\bibfnamefont {E.}~\bibnamefont {Demler}}, \bibinfo
  {author} {\bibfnamefont {S.}~\bibnamefont {Gopalakrishnan}}, \bibinfo
  {author} {\bibfnamefont {D.~A.}\ \bibnamefont {Huse}},\ and\ \bibinfo
  {author} {\bibfnamefont {M.}~\bibnamefont {Knap}},\ }\bibfield  {title}
  {\bibinfo {title} {Rare-region effects and dynamics near the many-body
  localization transition},\ }\href {https://doi.org/10.1002/andp.201600326}
  {\bibfield  {journal} {\bibinfo  {journal} {Ann. Phys.}\ }\textbf {\bibinfo
  {volume} {529}},\ \bibinfo {pages} {1600326} (\bibinfo {year}
  {2017})}\BibitemShut {NoStop}%
\bibitem [{\citenamefont {Luitz}\ and\ \citenamefont
  {Bar~Lev}(2017)}]{Luitz2016}%
  \BibitemOpen
  \bibfield  {author} {\bibinfo {author} {\bibfnamefont {D.~J.}\ \bibnamefont
  {Luitz}}\ and\ \bibinfo {author} {\bibfnamefont {Y.}~\bibnamefont
  {Bar~Lev}},\ }\bibfield  {title} {\bibinfo {title} {The ergodic side of the
  many-body localization transition},\ }\href
  {https://doi.org/10.1002/andp.201600350} {\bibfield  {journal} {\bibinfo
  {journal} {Ann. Phys.}\ }\textbf {\bibinfo {volume} {529}},\ \bibinfo {pages}
  {1600350} (\bibinfo {year} {2017})}\BibitemShut {NoStop}%
\bibitem [{\citenamefont {Swingle}\ \emph {et~al.}(2016)\citenamefont
  {Swingle}, \citenamefont {Bentsen}, \citenamefont {{Schleier-Smith}},\ and\
  \citenamefont {Hayden}}]{Swingle2016a}%
  \BibitemOpen
  \bibfield  {author} {\bibinfo {author} {\bibfnamefont {B.}~\bibnamefont
  {Swingle}}, \bibinfo {author} {\bibfnamefont {G.}~\bibnamefont {Bentsen}},
  \bibinfo {author} {\bibfnamefont {M.}~\bibnamefont {{Schleier-Smith}}},\ and\
  \bibinfo {author} {\bibfnamefont {P.}~\bibnamefont {Hayden}},\ }\bibfield
  {title} {\bibinfo {title} {Measuring the scrambling of quantum information},\
  }\href {https://doi.org/10.1103/PhysRevA.94.040302} {\bibfield  {journal}
  {\bibinfo  {journal} {Phys. Rev. A}\ }\textbf {\bibinfo {volume} {94}},\
  \bibinfo {pages} {040302} (\bibinfo {year} {2016})}\BibitemShut {NoStop}%
\bibitem [{\citenamefont {G{\"a}rttner}\ \emph {et~al.}(2017)\citenamefont
  {G{\"a}rttner}, \citenamefont {Bohnet}, \citenamefont {{Safavi-Naini}},
  \citenamefont {Wall}, \citenamefont {Bollinger},\ and\ \citenamefont
  {Rey}}]{Garttner2017}%
  \BibitemOpen
  \bibfield  {author} {\bibinfo {author} {\bibfnamefont {M.}~\bibnamefont
  {G{\"a}rttner}}, \bibinfo {author} {\bibfnamefont {J.~G.}\ \bibnamefont
  {Bohnet}}, \bibinfo {author} {\bibfnamefont {A.}~\bibnamefont
  {{Safavi-Naini}}}, \bibinfo {author} {\bibfnamefont {M.~L.}\ \bibnamefont
  {Wall}}, \bibinfo {author} {\bibfnamefont {J.~J.}\ \bibnamefont
  {Bollinger}},\ and\ \bibinfo {author} {\bibfnamefont {A.~M.}\ \bibnamefont
  {Rey}},\ }\bibfield  {title} {\bibinfo {title} {Measuring out-of-time-order
  correlations and multiple quantum spectra in a trapped-ion quantum magnet},\
  }\href {https://doi.org/10.1038/nphys4119} {\bibfield  {journal} {\bibinfo
  {journal} {Nat. Phys.}\ }\textbf {\bibinfo {volume} {13}},\ \bibinfo {pages}
  {781} (\bibinfo {year} {2017})}\BibitemShut {NoStop}%
\bibitem [{\citenamefont {Borgonovi}\ \emph {et~al.}(2019)\citenamefont
  {Borgonovi}, \citenamefont {Izrailev},\ and\ \citenamefont
  {Santos}}]{Borgonovi2019:exponentially}%
  \BibitemOpen
  \bibfield  {author} {\bibinfo {author} {\bibfnamefont {F.}~\bibnamefont
  {Borgonovi}}, \bibinfo {author} {\bibfnamefont {F.~M.}\ \bibnamefont
  {Izrailev}},\ and\ \bibinfo {author} {\bibfnamefont {L.~F.}\ \bibnamefont
  {Santos}},\ }\bibfield  {title} {\bibinfo {title} {Exponentially fast
  dynamics of chaotic many-body systems},\ }\href
  {https://doi.org/10.1103/physreve.99.010101} {\bibfield  {journal} {\bibinfo
  {journal} {Phys. Rev. E.}\ }\textbf {\bibinfo {volume} {99}},\ \bibinfo
  {pages} {010101} (\bibinfo {year} {2019})}\BibitemShut {NoStop}%
\bibitem [{\citenamefont {S{\'{a}}nchez}\ \emph {et~al.}(2020)\citenamefont
  {S{\'{a}}nchez}, \citenamefont {Chattah}, \citenamefont {Wei}, \citenamefont
  {Buljubasich}, \citenamefont {Cappellaro},\ and\ \citenamefont
  {Pastawski}}]{Sanchez2020}%
  \BibitemOpen
  \bibfield  {author} {\bibinfo {author} {\bibfnamefont {C.~M.}\ \bibnamefont
  {S{\'{a}}nchez}}, \bibinfo {author} {\bibfnamefont {A.~K.}\ \bibnamefont
  {Chattah}}, \bibinfo {author} {\bibfnamefont {K.~X.}\ \bibnamefont {Wei}},
  \bibinfo {author} {\bibfnamefont {L.}~\bibnamefont {Buljubasich}}, \bibinfo
  {author} {\bibfnamefont {P.}~\bibnamefont {Cappellaro}},\ and\ \bibinfo
  {author} {\bibfnamefont {H.~M.}\ \bibnamefont {Pastawski}},\ }\bibfield
  {title} {\bibinfo {title} {{Perturbation Independent Decay of the Loschmidt
  Echo in a Many-Body System}},\ }\href
  {https://doi.org/10.1103/PhysRevLett.124.030601} {\bibfield  {journal}
  {\bibinfo  {journal} {Phys. Rev. Lett.}\ }\textbf {\bibinfo {volume} {124}},\
  \bibinfo {pages} {30601} (\bibinfo {year} {2020})}\BibitemShut {NoStop}%
\bibitem [{\citenamefont {Niknam}\ \emph {et~al.}(2020)\citenamefont {Niknam},
  \citenamefont {Santos},\ and\ \citenamefont {Cory}}]{Niknam2020}%
  \BibitemOpen
  \bibfield  {author} {\bibinfo {author} {\bibfnamefont {M.}~\bibnamefont
  {Niknam}}, \bibinfo {author} {\bibfnamefont {L.~F.}\ \bibnamefont {Santos}},\
  and\ \bibinfo {author} {\bibfnamefont {D.~G.}\ \bibnamefont {Cory}},\
  }\bibfield  {title} {\bibinfo {title} {Sensitivity of quantum information to
  environment perturbations measured with a nonlocal out-of-time-order
  correlation function},\ }\href
  {https://doi.org/10.1103/PhysRevResearch.2.013200} {\bibfield  {journal}
  {\bibinfo  {journal} {Phys. Rev. Research}\ }\textbf {\bibinfo {volume}
  {2}},\ \bibinfo {pages} {013200} (\bibinfo {year} {2020})}\BibitemShut
  {NoStop}%
\bibitem [{\citenamefont {Guhr}\ \emph {et~al.}(1998)\citenamefont {Guhr},
  \citenamefont {{M{\"u}ller--Groeling}},\ and\ \citenamefont
  {Weidenm{\"u}ller}}]{Guhr1998}%
  \BibitemOpen
  \bibfield  {author} {\bibinfo {author} {\bibfnamefont {T.}~\bibnamefont
  {Guhr}}, \bibinfo {author} {\bibfnamefont {A.}~\bibnamefont
  {{M{\"u}ller--Groeling}}},\ and\ \bibinfo {author} {\bibfnamefont {H.~A.}\
  \bibnamefont {Weidenm{\"u}ller}},\ }\bibfield  {title} {\bibinfo {title}
  {Random-matrix theories in quantum physics: Common concepts},\ }\href
  {https://doi.org/10.1016/S0370-1573(97)00088-4} {\bibfield  {journal}
  {\bibinfo  {journal} {Phys. Rep.}\ }\textbf {\bibinfo {volume} {299}},\
  \bibinfo {pages} {189} (\bibinfo {year} {1998})}\BibitemShut {NoStop}%
\bibitem [{\citenamefont {Lozej}\ \emph {et~al.}(2021)\citenamefont {Lozej},
  \citenamefont {Casati},\ and\ \citenamefont {Prosen}}]{Lozej2021}%
  \BibitemOpen
  \bibfield  {author} {\bibinfo {author} {\bibfnamefont {{\v C}.}~\bibnamefont
  {Lozej}}, \bibinfo {author} {\bibfnamefont {G.}~\bibnamefont {Casati}},\ and\
  \bibinfo {author} {\bibfnamefont {T.}~\bibnamefont {Prosen}},\ }\href@noop {}
  {\bibinfo {title} {Quantum chaos in triangular billiards}} (\bibinfo {year}
  {2021}),\ \Eprint {https://arxiv.org/abs/2110.04168} {arXiv:2110.04168}
  \BibitemShut {NoStop}%
\bibitem [{\citenamefont {Wang}\ \emph {et~al.}(2022)\citenamefont {Wang},
  \citenamefont {Benenti}, \citenamefont {Casati},\ and\ \citenamefont
  {Wang}}]{Wang2022}%
  \BibitemOpen
  \bibfield  {author} {\bibinfo {author} {\bibfnamefont {J.}~\bibnamefont
  {Wang}}, \bibinfo {author} {\bibfnamefont {G.}~\bibnamefont {Benenti}},
  \bibinfo {author} {\bibfnamefont {G.}~\bibnamefont {Casati}},\ and\ \bibinfo
  {author} {\bibfnamefont {W.-G.}\ \bibnamefont {Wang}},\ }\href@noop {}
  {\bibinfo {title} {Statistical and dynamical properties of the quantum
  triangle map}} (\bibinfo {year} {2022}),\ \Eprint
  {https://arxiv.org/abs/2201.05921} {arXiv:2201.05921} \BibitemShut {NoStop}%
\bibitem [{\citenamefont {Akila}\ \emph {et~al.}(2017)\citenamefont {Akila},
  \citenamefont {Waltner}, \citenamefont {Gutkin}, \citenamefont {Braun},\ and\
  \citenamefont {Guhr}}]{PhysRevLett.118.164101}%
  \BibitemOpen
  \bibfield  {author} {\bibinfo {author} {\bibfnamefont {M.}~\bibnamefont
  {Akila}}, \bibinfo {author} {\bibfnamefont {D.}~\bibnamefont {Waltner}},
  \bibinfo {author} {\bibfnamefont {B.}~\bibnamefont {Gutkin}}, \bibinfo
  {author} {\bibfnamefont {P.}~\bibnamefont {Braun}},\ and\ \bibinfo {author}
  {\bibfnamefont {T.}~\bibnamefont {Guhr}},\ }\bibfield  {title} {\bibinfo
  {title} {{Semiclassical Identification of Periodic Orbits in a Quantum
  Many-Body System}},\ }\href {https://doi.org/10.1103/PhysRevLett.118.164101}
  {\bibfield  {journal} {\bibinfo  {journal} {Phys. Rev. Lett.}\ }\textbf
  {\bibinfo {volume} {118}},\ \bibinfo {pages} {164101} (\bibinfo {year}
  {2017})}\BibitemShut {NoStop}%
\bibitem [{\citenamefont {Avishai}\ \emph {et~al.}(2002)\citenamefont
  {Avishai}, \citenamefont {Richert},\ and\ \citenamefont
  {Berkovits}}]{Avishai2002}%
  \BibitemOpen
  \bibfield  {author} {\bibinfo {author} {\bibfnamefont {Y.}~\bibnamefont
  {Avishai}}, \bibinfo {author} {\bibfnamefont {J.}~\bibnamefont {Richert}},\
  and\ \bibinfo {author} {\bibfnamefont {R.}~\bibnamefont {Berkovits}},\
  }\bibfield  {title} {\bibinfo {title} {Level statistics in a {{Heisenberg}}
  chain with random magnetic field},\ }\href
  {https://doi.org/10.1103/PhysRevB.66.052416} {\bibfield  {journal} {\bibinfo
  {journal} {Phys. Rev. B}\ }\textbf {\bibinfo {volume} {66}},\ \bibinfo
  {pages} {052416} (\bibinfo {year} {2002})}\BibitemShut {NoStop}%
\bibitem [{\citenamefont {Richter}\ \emph {et~al.}(2019)\citenamefont
  {Richter}, \citenamefont {Casper}, \citenamefont {Brenig},\ and\
  \citenamefont {Steinigeweg}}]{Richter2019}%
  \BibitemOpen
  \bibfield  {author} {\bibinfo {author} {\bibfnamefont {J.}~\bibnamefont
  {Richter}}, \bibinfo {author} {\bibfnamefont {N.}~\bibnamefont {Casper}},
  \bibinfo {author} {\bibfnamefont {W.}~\bibnamefont {Brenig}},\ and\ \bibinfo
  {author} {\bibfnamefont {R.}~\bibnamefont {Steinigeweg}},\ }\bibfield
  {title} {\bibinfo {title} {Magnetization dynamics in clean and disordered
  spin-1 {XXZ} chains},\ }\href {https://doi.org/10.1103/physrevb.100.144423}
  {\bibfield  {journal} {\bibinfo  {journal} {Phys. Rev. B}\ }\textbf {\bibinfo
  {volume} {100}},\ \bibinfo {pages} {144423} (\bibinfo {year}
  {2019})}\BibitemShut {NoStop}%
\bibitem [{\citenamefont {Santos}\ \emph {et~al.}(2020)\citenamefont {Santos},
  \citenamefont {P{\'{e}}rez-Bernal},\ and\ \citenamefont
  {Torres-Herrera}}]{Santos2020}%
  \BibitemOpen
  \bibfield  {author} {\bibinfo {author} {\bibfnamefont {L.~F.}\ \bibnamefont
  {Santos}}, \bibinfo {author} {\bibfnamefont {F.}~\bibnamefont
  {P{\'{e}}rez-Bernal}},\ and\ \bibinfo {author} {\bibfnamefont {E.~J.}\
  \bibnamefont {Torres-Herrera}},\ }\bibfield  {title} {\bibinfo {title} {Speck
  of chaos},\ }\href {https://doi.org/10.1103/physrevresearch.2.043034}
  {\bibfield  {journal} {\bibinfo  {journal} {Phys. Rev. Research}\ }\textbf
  {\bibinfo {volume} {2}},\ \bibinfo {pages} {043034} (\bibinfo {year}
  {2020})}\BibitemShut {NoStop}%
\bibitem [{\citenamefont {de~Wijn}\ \emph {et~al.}(2012)\citenamefont
  {de~Wijn}, \citenamefont {Hess},\ and\ \citenamefont {Fine}}]{deWijn2012}%
  \BibitemOpen
  \bibfield  {author} {\bibinfo {author} {\bibfnamefont {A.~S.}\ \bibnamefont
  {de~Wijn}}, \bibinfo {author} {\bibfnamefont {B.}~\bibnamefont {Hess}},\ and\
  \bibinfo {author} {\bibfnamefont {B.~V.}\ \bibnamefont {Fine}},\ }\bibfield
  {title} {\bibinfo {title} {Largest lyapunov exponents for lattices of
  interacting classical spins},\ }\href
  {https://doi.org/10.1103/physrevlett.109.034101} {\bibfield  {journal}
  {\bibinfo  {journal} {Phys. Rev. Lett.}\ }\textbf {\bibinfo {volume} {109}},\
  \bibinfo {pages} {034101} (\bibinfo {year} {2012})}\BibitemShut {NoStop}%
\bibitem [{\citenamefont {Elsayed}\ and\ \citenamefont
  {Fine}(2015)}]{Elsayed2015}%
  \BibitemOpen
  \bibfield  {author} {\bibinfo {author} {\bibfnamefont {T.~A.}\ \bibnamefont
  {Elsayed}}\ and\ \bibinfo {author} {\bibfnamefont {B.~V.}\ \bibnamefont
  {Fine}},\ }\bibfield  {title} {\bibinfo {title} {Sensitivity to small
  perturbations in systems of large quantum spins},\ }\href
  {https://doi.org/10.1088/0031-8949/2015/t165/014011} {\bibfield  {journal}
  {\bibinfo  {journal} {Phys. Scripta}\ }\textbf {\bibinfo {volume} {T165}},\
  \bibinfo {pages} {014011} (\bibinfo {year} {2015})}\BibitemShut {NoStop}%
\bibitem [{\citenamefont {Schiulaz}\ \emph {et~al.}(2018)\citenamefont
  {Schiulaz}, \citenamefont {T{\'a}vora},\ and\ \citenamefont
  {Santos}}]{schiulaz2018:FewManybodyQuantum}%
  \BibitemOpen
  \bibfield  {author} {\bibinfo {author} {\bibfnamefont {M.}~\bibnamefont
  {Schiulaz}}, \bibinfo {author} {\bibfnamefont {M.}~\bibnamefont
  {T{\'a}vora}},\ and\ \bibinfo {author} {\bibfnamefont {L.~F.}\ \bibnamefont
  {Santos}},\ }\bibfield  {title} {\bibinfo {title} {From few- to many-body
  quantum systems},\ }\href {https://doi.org/10.1088/2058-9565/aad913}
  {\bibfield  {journal} {\bibinfo  {journal} {Quantum Sci. Technol.}\ }\textbf
  {\bibinfo {volume} {3}},\ \bibinfo {pages} {044006} (\bibinfo {year}
  {2018})}\BibitemShut {NoStop}%
\bibitem [{\citenamefont {Zisling}\ \emph {et~al.}(2021)\citenamefont
  {Zisling}, \citenamefont {Santos},\ and\ \citenamefont
  {Bar~Lev}}]{Zisling2021}%
  \BibitemOpen
  \bibfield  {author} {\bibinfo {author} {\bibfnamefont {G.}~\bibnamefont
  {Zisling}}, \bibinfo {author} {\bibfnamefont {L.}~\bibnamefont {Santos}},\
  and\ \bibinfo {author} {\bibfnamefont {Y.}~\bibnamefont {Bar~Lev}},\
  }\bibfield  {title} {\bibinfo {title} {How many particles make up a chaotic
  many-body quantum system?},\ }\href
  {https://doi.org/10.21468/SciPostPhys.10.4.088} {\bibfield  {journal}
  {\bibinfo  {journal} {SciPost Phys.}\ }\textbf {\bibinfo {volume} {10}},\
  \bibinfo {pages} {088} (\bibinfo {year} {2021})}\BibitemShut {NoStop}%
\bibitem [{\citenamefont {Mirkin}\ and\ \citenamefont
  {Wisniacki}(2021)}]{Mirkin2021:QuantumChaos}%
  \BibitemOpen
  \bibfield  {author} {\bibinfo {author} {\bibfnamefont {N.}~\bibnamefont
  {Mirkin}}\ and\ \bibinfo {author} {\bibfnamefont {D.}~\bibnamefont
  {Wisniacki}},\ }\bibfield  {title} {\bibinfo {title} {Quantum chaos,
  equilibration, and control in extremely short spin chains},\ }\href
  {https://doi.org/10.1103/PhysRevE.103.L020201} {\bibfield  {journal}
  {\bibinfo  {journal} {Phys. Rev. E}\ }\textbf {\bibinfo {volume} {103}},\
  \bibinfo {pages} {L020201} (\bibinfo {year} {2021})}\BibitemShut {NoStop}%
\bibitem [{\citenamefont {Santos}\ \emph {et~al.}(2004)\citenamefont {Santos},
  \citenamefont {Rigolin},\ and\ \citenamefont
  {Escobar}}]{Santos2004:Entangelement}%
  \BibitemOpen
  \bibfield  {author} {\bibinfo {author} {\bibfnamefont {L.~F.}\ \bibnamefont
  {Santos}}, \bibinfo {author} {\bibfnamefont {G.}~\bibnamefont {Rigolin}},\
  and\ \bibinfo {author} {\bibfnamefont {C.~O.}\ \bibnamefont {Escobar}},\
  }\bibfield  {title} {\bibinfo {title} {Entanglement versus chaos in
  disordered spin chains},\ }\href {https://doi.org/10.1103/physreva.69.042304}
  {\bibfield  {journal} {\bibinfo  {journal} {Phys. Rev. A}\ }\textbf {\bibinfo
  {volume} {69}},\ \bibinfo {pages} {042304} (\bibinfo {year}
  {2004})}\BibitemShut {NoStop}%
\bibitem [{\citenamefont {Santos}(2004)}]{Santos2004i:Integrability}%
  \BibitemOpen
  \bibfield  {author} {\bibinfo {author} {\bibfnamefont {L.~F.}\ \bibnamefont
  {Santos}},\ }\bibfield  {title} {\bibinfo {title} {Integrability of a
  disordered heisenberg spin-1/2 chain},\ }\href
  {https://doi.org/10.1088/0305-4470/37/17/004} {\bibfield  {journal} {\bibinfo
   {journal} {J. Phys. A}\ }\textbf {\bibinfo {volume} {37}},\ \bibinfo {pages}
  {4723} (\bibinfo {year} {2004})}\BibitemShut {NoStop}%
\bibitem [{\citenamefont {Elsayed}\ \emph {et~al.}(2014)\citenamefont
  {Elsayed}, \citenamefont {Hess},\ and\ \citenamefont {Fine}}]{Elsayed2014}%
  \BibitemOpen
  \bibfield  {author} {\bibinfo {author} {\bibfnamefont {T.~A.}\ \bibnamefont
  {Elsayed}}, \bibinfo {author} {\bibfnamefont {B.}~\bibnamefont {Hess}},\ and\
  \bibinfo {author} {\bibfnamefont {B.~V.}\ \bibnamefont {Fine}},\ }\bibfield
  {title} {\bibinfo {title} {{Signatures of chaos in time series generated by
  many-spin systems at high temperatures}},\ }\href
  {https://doi.org/10.1103/PhysRevE.90.022910} {\bibfield  {journal} {\bibinfo
  {journal} {Phys. Rev. E}\ }\textbf {\bibinfo {volume} {90}},\ \bibinfo
  {pages} {022910} (\bibinfo {year} {2014})}\BibitemShut {NoStop}%
\bibitem [{\citenamefont {Petzold}(1980)}]{Petzold1980}%
  \BibitemOpen
  \bibfield  {author} {\bibinfo {author} {\bibfnamefont {L.}~\bibnamefont
  {Petzold}},\ }\href@noop {} {{\selectlanguage {English}\emph {\bibinfo
  {title} {{Automatic selection of methods for solving stiff and nonstiff
  systems of ordinary differential equations}}}}},\ \bibinfo {type} {Tech.
  Rep.}\ (\bibinfo {address} {United States},\ \bibinfo {year}
  {1980})\BibitemShut {NoStop}%
\bibitem [{\citenamefont {Pechukas}(1983)}]{Pechukas1983}%
  \BibitemOpen
  \bibfield  {author} {\bibinfo {author} {\bibfnamefont {P.}~\bibnamefont
  {Pechukas}},\ }\bibfield  {title} {\bibinfo {title} {Distribution of {{Energy
  Eigenvalues}} in the {{Irregular Spectrum}}},\ }\href
  {https://doi.org/10.1103/PhysRevLett.51.943} {\bibfield  {journal} {\bibinfo
  {journal} {Phys. Rev. Lett.}\ }\textbf {\bibinfo {volume} {51}},\ \bibinfo
  {pages} {943} (\bibinfo {year} {1983})}\BibitemShut {NoStop}%
\bibitem [{\citenamefont {Pechukas}(1984)}]{Pechukas1984}%
  \BibitemOpen
  \bibfield  {author} {\bibinfo {author} {\bibfnamefont {P.}~\bibnamefont
  {Pechukas}},\ }\bibfield  {title} {\bibinfo {title} {Remarks on "quantum
  chaos"},\ }\href {https://doi.org/10.1021/j150665a006} {\bibfield  {journal}
  {\bibinfo  {journal} {J. Phys. Chem.}\ }\textbf {\bibinfo {volume} {88}},\
  \bibinfo {pages} {4823} (\bibinfo {year} {1984})}\BibitemShut {NoStop}%
\bibitem [{\citenamefont {Peres}(1984{\natexlab{a}})}]{Peres1984}%
  \BibitemOpen
  \bibfield  {author} {\bibinfo {author} {\bibfnamefont {A.}~\bibnamefont
  {Peres}},\ }\bibfield  {title} {\bibinfo {title} {Ergodicity and mixing in
  quantum theory. {{I}}},\ }\href {https://doi.org/10.1103/PhysRevA.30.504}
  {\bibfield  {journal} {\bibinfo  {journal} {Phys. Rev. A}\ }\textbf {\bibinfo
  {volume} {30}},\ \bibinfo {pages} {504} (\bibinfo {year}
  {1984}{\natexlab{a}})}\BibitemShut {NoStop}%
\bibitem [{\citenamefont {Peres}(1984{\natexlab{b}})}]{Peres1984a}%
  \BibitemOpen
  \bibfield  {author} {\bibinfo {author} {\bibfnamefont {A.}~\bibnamefont
  {Peres}},\ }\bibfield  {title} {\bibinfo {title} {Stability of quantum motion
  in chaotic and regular systems},\ }\href
  {https://doi.org/10.1103/PhysRevA.30.1610} {\bibfield  {journal} {\bibinfo
  {journal} {Phys. Rev. A}\ }\textbf {\bibinfo {volume} {30}},\ \bibinfo
  {pages} {1610} (\bibinfo {year} {1984}{\natexlab{b}})}\BibitemShut {NoStop}%
\bibitem [{\citenamefont {Feingold}\ \emph {et~al.}(1984)\citenamefont
  {Feingold}, \citenamefont {Moiseyev},\ and\ \citenamefont
  {Peres}}]{Feingold1984}%
  \BibitemOpen
  \bibfield  {author} {\bibinfo {author} {\bibfnamefont {M.}~\bibnamefont
  {Feingold}}, \bibinfo {author} {\bibfnamefont {N.}~\bibnamefont {Moiseyev}},\
  and\ \bibinfo {author} {\bibfnamefont {A.}~\bibnamefont {Peres}},\ }\bibfield
   {title} {\bibinfo {title} {Ergodicity and mixing in quantum theory.
  {{II}}},\ }\href {https://doi.org/10.1103/PhysRevA.30.509} {\bibfield
  {journal} {\bibinfo  {journal} {Phys. Rev. A}\ }\textbf {\bibinfo {volume}
  {30}},\ \bibinfo {pages} {509} (\bibinfo {year} {1984})}\BibitemShut
  {NoStop}%
\bibitem [{\citenamefont {Feingold}\ \emph {et~al.}(1985)\citenamefont
  {Feingold}, \citenamefont {Moiseyev},\ and\ \citenamefont
  {Peres}}]{Feingold1985}%
  \BibitemOpen
  \bibfield  {author} {\bibinfo {author} {\bibfnamefont {M.}~\bibnamefont
  {Feingold}}, \bibinfo {author} {\bibfnamefont {N.}~\bibnamefont {Moiseyev}},\
  and\ \bibinfo {author} {\bibfnamefont {A.}~\bibnamefont {Peres}},\ }\bibfield
   {title} {\bibinfo {title} {Classical limit of quantum chaos},\ }\href
  {https://doi.org/10.1016/0009-2614(85)85241-6} {\bibfield  {journal}
  {\bibinfo  {journal} {Chem. Phys. Lett.}\ }\textbf {\bibinfo {volume}
  {117}},\ \bibinfo {pages} {344} (\bibinfo {year} {1985})}\BibitemShut
  {NoStop}%
\bibitem [{\citenamefont {Feingold}\ and\ \citenamefont
  {Peres}(1986)}]{Feingold1986}%
  \BibitemOpen
  \bibfield  {author} {\bibinfo {author} {\bibfnamefont {M.}~\bibnamefont
  {Feingold}}\ and\ \bibinfo {author} {\bibfnamefont {A.}~\bibnamefont
  {Peres}},\ }\bibfield  {title} {\bibinfo {title} {Distribution of matrix
  elements of chaotic systems},\ }\href
  {https://doi.org/10.1103/PhysRevA.34.591} {\bibfield  {journal} {\bibinfo
  {journal} {Phys. Rev. A}\ }\textbf {\bibinfo {volume} {34}},\ \bibinfo
  {pages} {591} (\bibinfo {year} {1986})}\BibitemShut {NoStop}%
\bibitem [{\citenamefont {Deutsch}(1991)}]{Deutsch1991}%
  \BibitemOpen
  \bibfield  {author} {\bibinfo {author} {\bibfnamefont {J.~M.}\ \bibnamefont
  {Deutsch}},\ }\bibfield  {title} {\bibinfo {title} {Quantum statistical
  mechanics in a closed system},\ }\href
  {https://doi.org/10.1103/PhysRevA.43.2046} {\bibfield  {journal} {\bibinfo
  {journal} {Phys. Rev. A}\ }\textbf {\bibinfo {volume} {43}},\ \bibinfo
  {pages} {2046} (\bibinfo {year} {1991})}\BibitemShut {NoStop}%
\bibitem [{\citenamefont {Srednicki}(1994)}]{Srednicki1994}%
  \BibitemOpen
  \bibfield  {author} {\bibinfo {author} {\bibfnamefont {M.}~\bibnamefont
  {Srednicki}},\ }\bibfield  {title} {\bibinfo {title} {Chaos and quantum
  thermalization},\ }\href {https://doi.org/10.1103/PhysRevE.50.888} {\bibfield
   {journal} {\bibinfo  {journal} {Phys. Rev. E}\ }\textbf {\bibinfo {volume}
  {50}},\ \bibinfo {pages} {888} (\bibinfo {year} {1994})}\BibitemShut
  {NoStop}%
\bibitem [{\citenamefont {Srednicki}(1996)}]{Srednicki1995}%
  \BibitemOpen
  \bibfield  {author} {\bibinfo {author} {\bibfnamefont {M.}~\bibnamefont
  {Srednicki}},\ }\bibfield  {title} {\bibinfo {title} {Thermal fluctuations in
  quantized chaotic systems},\ }\href
  {https://doi.org/10.1088/0305-4470/29/4/003} {\bibfield  {journal} {\bibinfo
  {journal} {J. Phys. Math. Gen.}\ }\textbf {\bibinfo {volume} {29}},\ \bibinfo
  {pages} {L75} (\bibinfo {year} {1996})}\BibitemShut {NoStop}%
\bibitem [{\citenamefont {Srednicki}(1999)}]{Srednicki1999}%
  \BibitemOpen
  \bibfield  {author} {\bibinfo {author} {\bibfnamefont {M.}~\bibnamefont
  {Srednicki}},\ }\bibfield  {title} {\bibinfo {title} {The approach to thermal
  equilibrium in quantized chaotic systems},\ }\href
  {https://doi.org/10.1088/0305-4470/32/7/007} {\bibfield  {journal} {\bibinfo
  {journal} {J. Phys. Math. Gen.}\ }\textbf {\bibinfo {volume} {32}},\ \bibinfo
  {pages} {1163} (\bibinfo {year} {1999})}\BibitemShut {NoStop}%
\bibitem [{\citenamefont {Rigol}\ \emph {et~al.}(2008)\citenamefont {Rigol},
  \citenamefont {Dunjko},\ and\ \citenamefont {Olshanii}}]{Rigol2008}%
  \BibitemOpen
  \bibfield  {author} {\bibinfo {author} {\bibfnamefont {M.}~\bibnamefont
  {Rigol}}, \bibinfo {author} {\bibfnamefont {V.}~\bibnamefont {Dunjko}},\ and\
  \bibinfo {author} {\bibfnamefont {M.}~\bibnamefont {Olshanii}},\ }\bibfield
  {title} {\bibinfo {title} {Thermalization and its mechanism for generic
  isolated quantum systems.},\ }\href {https://doi.org/10.1038/nature06838}
  {\bibfield  {journal} {\bibinfo  {journal} {Nature}\ }\textbf {\bibinfo
  {volume} {452}},\ \bibinfo {pages} {854} (\bibinfo {year}
  {2008})}\BibitemShut {NoStop}%
\bibitem [{\citenamefont {Beugeling}\ \emph {et~al.}(2015)\citenamefont
  {Beugeling}, \citenamefont {Moessner},\ and\ \citenamefont
  {Haque}}]{Beugeling2015}%
  \BibitemOpen
  \bibfield  {author} {\bibinfo {author} {\bibfnamefont {W.}~\bibnamefont
  {Beugeling}}, \bibinfo {author} {\bibfnamefont {R.}~\bibnamefont
  {Moessner}},\ and\ \bibinfo {author} {\bibfnamefont {M.}~\bibnamefont
  {Haque}},\ }\bibfield  {title} {\bibinfo {title} {Off-diagonal matrix
  elements of local operators in many-body quantum systems},\ }\href
  {https://doi.org/10.1103/PhysRevE.91.012144} {\bibfield  {journal} {\bibinfo
  {journal} {Phys. Rev. E}\ }\textbf {\bibinfo {volume} {91}},\ \bibinfo
  {pages} {012144} (\bibinfo {year} {2015})}\BibitemShut {NoStop}%
\bibitem [{\citenamefont {Leblond}\ \emph {et~al.}(2019)\citenamefont
  {Leblond}, \citenamefont {Mallayya}, \citenamefont {Vidmar},\ and\
  \citenamefont {Rigol}}]{Leblond2019}%
  \BibitemOpen
  \bibfield  {author} {\bibinfo {author} {\bibfnamefont {T.}~\bibnamefont
  {Leblond}}, \bibinfo {author} {\bibfnamefont {K.}~\bibnamefont {Mallayya}},
  \bibinfo {author} {\bibfnamefont {L.}~\bibnamefont {Vidmar}},\ and\ \bibinfo
  {author} {\bibfnamefont {M.}~\bibnamefont {Rigol}},\ }\bibfield  {title}
  {\bibinfo {title} {Entanglement and matrix elements of observables in
  interacting integrable systems},\ }\href
  {https://doi.org/10.1103/PhysRevE.100.062134} {\bibfield  {journal} {\bibinfo
   {journal} {Phys. Rev. E}\ }\textbf {\bibinfo {volume} {100}},\ \bibinfo
  {pages} {062134} (\bibinfo {year} {2019})}\BibitemShut {NoStop}%
\bibitem [{\citenamefont {Khaymovich}\ \emph {et~al.}(2019)\citenamefont
  {Khaymovich}, \citenamefont {Haque},\ and\ \citenamefont
  {McClarty}}]{Khaymovich2019}%
  \BibitemOpen
  \bibfield  {author} {\bibinfo {author} {\bibfnamefont {I.~M.}\ \bibnamefont
  {Khaymovich}}, \bibinfo {author} {\bibfnamefont {M.}~\bibnamefont {Haque}},\
  and\ \bibinfo {author} {\bibfnamefont {P.~A.}\ \bibnamefont {McClarty}},\
  }\bibfield  {title} {\bibinfo {title} {Eigenstate thermalization, random
  matrix theory, and behemoths},\ }\href
  {https://doi.org/10.1103/physrevlett.122.070601} {\bibfield  {journal}
  {\bibinfo  {journal} {Phys. Rev. Lett.}\ }\textbf {\bibinfo {volume} {122}},\
  \bibinfo {pages} {070601} (\bibinfo {year} {2019})}\BibitemShut {NoStop}%
\bibitem [{\citenamefont {{\L}yd{\.{z}}ba}\ \emph {et~al.}(2021)\citenamefont
  {{\L}yd{\.{z}}ba}, \citenamefont {Zhang}, \citenamefont {Rigol},\ and\
  \citenamefont {Vidmar}}]{ydba2021}%
  \BibitemOpen
  \bibfield  {author} {\bibinfo {author} {\bibfnamefont {P.}~\bibnamefont
  {{\L}yd{\.{z}}ba}}, \bibinfo {author} {\bibfnamefont {Y.}~\bibnamefont
  {Zhang}}, \bibinfo {author} {\bibfnamefont {M.}~\bibnamefont {Rigol}},\ and\
  \bibinfo {author} {\bibfnamefont {L.}~\bibnamefont {Vidmar}},\ }\bibfield
  {title} {\bibinfo {title} {Single-particle eigenstate thermalization in
  quantum-chaotic quadratic hamiltonians},\ }\href
  {https://doi.org/10.1103/physrevb.104.214203} {\bibfield  {journal} {\bibinfo
   {journal} {Phys. Rev. B}\ }\textbf {\bibinfo {volume} {104}},\ \bibinfo
  {pages} {214203} (\bibinfo {year} {2021})}\BibitemShut {NoStop}%
\bibitem [{\citenamefont {Oganesyan}\ and\ \citenamefont
  {Huse}(2007)}]{Oganesyan2007}%
  \BibitemOpen
  \bibfield  {author} {\bibinfo {author} {\bibfnamefont {V.}~\bibnamefont
  {Oganesyan}}\ and\ \bibinfo {author} {\bibfnamefont {D.~A.}\ \bibnamefont
  {Huse}},\ }\bibfield  {title} {\bibinfo {title} {Localization of interacting
  fermions at high temperature},\ }\href
  {https://doi.org/10.1103/PhysRevB.75.155111} {\bibfield  {journal} {\bibinfo
  {journal} {Phys. Rev. B}\ }\textbf {\bibinfo {volume} {75}},\ \bibinfo
  {pages} {155111} (\bibinfo {year} {2007})}\BibitemShut {NoStop}%
\bibitem [{\citenamefont {Atas}\ \emph {et~al.}(2013)\citenamefont {Atas},
  \citenamefont {Bogomolny}, \citenamefont {Giraud},\ and\ \citenamefont
  {Roux}}]{Atas2013}%
  \BibitemOpen
  \bibfield  {author} {\bibinfo {author} {\bibfnamefont {Y.~Y.}\ \bibnamefont
  {Atas}}, \bibinfo {author} {\bibfnamefont {E.}~\bibnamefont {Bogomolny}},
  \bibinfo {author} {\bibfnamefont {O.}~\bibnamefont {Giraud}},\ and\ \bibinfo
  {author} {\bibfnamefont {G.}~\bibnamefont {Roux}},\ }\bibfield  {title}
  {\bibinfo {title} {Distribution of the {{Ratio}} of {{Consecutive Level
  Spacings}} in {{Random Matrix Ensembles}}},\ }\href
  {https://doi.org/10.1103/PhysRevLett.110.084101} {\bibfield  {journal}
  {\bibinfo  {journal} {Phys. Rev. Lett.}\ }\textbf {\bibinfo {volume} {110}},\
  \bibinfo {pages} {084101} (\bibinfo {year} {2013})}\BibitemShut {NoStop}%
\bibitem [{\citenamefont {Corps}\ and\ \citenamefont
  {Rela{\~{n}}o}(2020)}]{Corps2020}%
  \BibitemOpen
  \bibfield  {author} {\bibinfo {author} {\bibfnamefont {{\'{A}}.~L.}\
  \bibnamefont {Corps}}\ and\ \bibinfo {author} {\bibfnamefont
  {A.}~\bibnamefont {Rela{\~{n}}o}},\ }\bibfield  {title} {\bibinfo {title}
  {{Distribution of the ratio of consecutive level spacings for different
  symmetries and degrees of chaos}},\ }\href
  {https://doi.org/10.1103/PhysRevE.101.022222} {\bibfield  {journal} {\bibinfo
   {journal} {Phys. Rev. E}\ }\textbf {\bibinfo {volume} {101}},\ \bibinfo
  {pages} {022222} (\bibinfo {year} {2020})}\BibitemShut {NoStop}%
\end{thebibliography}%

\end{document}